\documentclass[12pt]{article}
\usepackage{amssymb}

\newcommand{\be}{\begin{equation}}
\newcommand{\ee}{\end{equation}}

\newcommand{\cP}{{\cal P}}

\begin{document}

\begin{center}
{\large \bf pQCD phenomenology of elastic $ed$ scattering }
\end{center}
\begin{center} A.P.~Kobushkin $^{a),\ b)}$ and
Ya.D.~Krivenko $^{c)}$\\
$^{a)}${\it N N Bogolyubov Institute for Theoretical Physics}\\
{\it Metrologicheskaya str. 14B, Kiev, 03143, Ukraine}\\
$^{b)}${\it Research Center for Nuclear Physics, Osaka University (Suita Campus)}\\
{\it 10-1, Mihogaoka Ibaraki Osaka 567-0047, Japan}\\
$^{c)}${\it Institute for Nuclear Research}\\
{\it Prospekt Nauki 47,Kiev, 03028, Ukraine}
\end{center}

\vspace{.5cm}
\begin{abstract}
Electron-deuteron elastic scattering data ($A(Q^2)$ and $B(Q^2)$ structure
functions and polarization observables $t_{20}$,  $t_{21}$ and $t_{22}$)
are fit with a model that respects asymptotic properties of pQCD at high 
momentum transfer. The data analysis shows that pQCD 
starts from $Q^2_{\rm QCD}=3.5\ \rm (GeV/c)^2$. Predictions for the magnetic structure 
function  $B(Q^2)$ and the polarization observables at high momentum 
transfer are given.\\[0.5 cm]
{\bf PACS} number(s): 13.40.Fn, 12.38.Bx, 25.30.Bf
\end{abstract}

\section{Introduction}
In recent few years new data from TJINAF on the $ed$ elastic scattering were 
reported. They include the electric structure function, $A(Q^2)$, measured 
with high precision up to $Q^2=6\ \rm{(GeV/c)^2}$ \cite{Alexa,AbbottA} and
measurements of 
tensor polarization observables, $t_{20}$, $t_{21}$ and $t_{22}$, up to 
$Q^2=1.7\ \rm{(GeV/c)^2}$ \cite{Abbottt20}. 

This data, together with data on the magnetic structure function,
 $B(Q^2)$ \cite{Bosted}, restrict the deuteron structure at scales where 
quark-gluon degrees of freedom are expected to become defrozen.
For example, according to optimistic estimations pQCD should start from $Q^2$
of order of few $\rm (GeV/c)^2$ \cite{BJL}. It is nice that this 
prediction was confirmed by analysis of TJINAF data on $A(Q^2)$ at   
$Q^2 > 2\ \rm{(GeV/c)^2}$ \cite{AbbottA}.

For further conclusions one also should consider the spin structure of the 
deuteron from pQCD point of view. However data on polarization observables, 
as well as on $B(Q^2)$, correspond to $Q^2 \lesssim 2\ \rm{(GeV/c)^2}$, which is 
not enough for pQCD. This is a typical intermediate region between nucleon-meson 
and quark-gluon pictures, where isobar configurations, meson exchange currents 
and constituent quark degrees of freedom are all important \cite{Lomon}.  

The purpose of this work is to investigate phenomenologically a smooth
connection between nucleon-meson and pQCD regions and make predictions for
$B(Q^2)$ and the polarization observables at higher $Q^2$, where pQCD should work.
A parameterization which connects these two regions was proposed earlier
by one of the authors (A.P.~K.) and A.I.~Syamtomov \cite{KS2}. 
It assumes power fall off of helicity spin amplitudes at asymptotically high $Q^2$
coming from quark counting rules. A new analysis of the parameterization
\cite{KS2} which includes the recent TJINAF data  was provided in \cite{Abbott_par}.
Now we study logarithmic corrections to the power behavior.
Such corrections are shown to be important for the structure function
$A(Q^2)$ at the highest region of TJINAF energy \cite{AbbottA}.

The paper is organized as follows. In sect.~2 we discuss the general structure 
of the helicity amplitudes for the elastic $ed$ scattering in the light cone frame (LCF)
and pQCD predictions for the helicity amplitudes at high $Q^2$.
Parameterization of the helicity amplitudes which smoothly connects regions of low 
and high $Q^2$ is given in sect.~3. Then, in sect.~4, the data base and fitting
procedure are summarized. Discussions and summary are given in sect.~5.

\section{Helicity amplitudes and the deuteron form factors}
\subsection{Helicity amplitudes in LCF}
The main object of our analysis is the helicity amplitudes of the
$\gamma^{\ast} +d\to d$ transition
\begin{equation}
J^{\mu}_{\lambda^{'} \lambda}=\langle
p^{'},\lambda^{'}|j^{\mu}|p,\lambda \rangle,
\label{1}
\end{equation}
where $p$ and $p'$ are momenta and $\lambda$ and $\lambda '$ are
helicities of the deuteron in the initial and final states, respectively.

Due to gauge invariance, covariance and discrete symmetries only three
of the 36 helicity amplitudes (\ref{1}) are independent and one
can choose different sets of independent helicity amplitudes. Direct 
calculations, however, demonstrate that it is not so in dynamics at 
LCF \cite{GrachKondratyuk}. This phenomena was shown to come from the 
incompatibility of transformation properties of approximate current 
and a deuteron wave function used in practical calculations 
\cite{KarmanovSmirnov,Karmanov}. 
As a result a non-physical dependence on orientation of light-front plane 
appears. Thus the choice of the independent amplitudes becomes of great 
importance in pQCD calculations where LCF is often used.

First let us define LCF as a limiting reference system where the 
$z$-projection of the incoming and outgoing deuteron is close to infinity. In
LCF the momenta of the incoming and outgoing deuteron are given as follows 
\begin{eqnarray}
p^{\mu}&=&\left({\cP}+\frac{M^2+p^2_{\perp}}{4\cP},\vec p_{\perp},
\cP-\frac{M^2+p^2_{\perp}}{4\cP}\right)  ,\nonumber \\
p'^{\mu}&=&\left({\cP}+\frac{M^2+p^2_{\perp}}{4\cP},-\vec p_{\perp},
\cP-\frac{M^2+p^2_{\perp}}{4\cP}\right)
\label{d_mom}
\end{eqnarray}
with
\begin{equation}
p^{+}\equiv p^{0} + p^{3}=2\cP,\
p'^{\,+}\equiv p'^{\,0} + p'^{\,3}=2\cP,\ \cP\gg M^2+ p\,^2_{\perp}
\label{3}
\end{equation}
($M$ is the deuteron mass). The momentum of the virtual photon  
is given by
\begin{equation}
q^{\mu}=(0,-2\vec p_{\perp},0),\ \vec
p_{\perp}=-\left(\frac{1}{2}Q,0\right)
\label{photon_mom}
\end{equation}
and the polarization vectors for the deuteron in the initial and final states, 
respectively, read
\begin{eqnarray}
\varepsilon^{\mu}(\lambda=\pm 1, p)&=&
-\sqrt{\frac{1}{2}}
\left(\pm \frac{p_{\perp}}{2\cP}, \pm 1, i, \mp \frac{p_{\perp}}
{2\cP}\right) ,\nonumber \\
\varepsilon^{\mu}(\lambda=0, p)&=&
\frac{1}{M}\left(\cP-\frac{M^2-p^2_{\perp}}{4\cP},
\vec p_{\perp},\cP+\frac{M^2-p^2_{\perp}}{4\cP}\right), \nonumber \\
\varepsilon^{\mu}(\lambda'=\pm 1, p')&=&
-\sqrt{\frac{1}{2}}\left(\mp \frac{p_{\perp}}{2\cP},
\pm 1, i, \pm \frac{p_{\perp}}{2\cP}\right),\nonumber \\
\varepsilon^{\mu}(\lambda=0, p')&=&
\frac{1}{M}\left(\cP-\frac{M^2-p^2_{\perp}}{4\cP},
-\vec p_{\perp},\cP+\frac{M^2-p^2_{\perp}}{4\cP}\right).
\label{polarization}
\end{eqnarray}
Here we put $p_{\perp}\equiv \sqrt{\vec p^{\, 2}_{\perp}}$.
Using the standard expression for the e.m. current matrix element
\begin{eqnarray}
J_{\lambda'\lambda}^{\mu}=
&-&\left\{G_1(Q^2)\varepsilon^{\ast}(\lambda',p')\varepsilon(\lambda,p)
(p+p')^{\mu}+\right.\nonumber \\
&+&\left.G_2(Q^2)[\varepsilon^{\mu}(\lambda,p)
\left(\varepsilon^{\ast}(\lambda',p')q\right)-
\varepsilon^{\ast\mu}(\lambda',p')
\left(\varepsilon(\lambda,p)q\right)]-\right.\nonumber\\
&-&\left.G_3(Q^2)(p+p')^{\mu}
\frac{\left(\varepsilon^{\ast}(\lambda',p')q\right)
\left(\varepsilon(\lambda,p)q\right)}{2M^2}
\right\}
\label{current_std}
\end{eqnarray}
one gets the following expressions for the current plus-component  
\begin{eqnarray}
J_{00}^{+}&=&p^{+}\left\{
2(1-2\eta)G_1+4\eta G_2-4\eta^2G_3\right\},
\label{00} \\
J_{10}^{+}&=&p^{+} \left\{
2\sqrt{2\eta}G_1-\sqrt{2\eta}G_2+2\sqrt{2\eta}\eta G_3 \right\},
\label{10} \\
J_{1-1}^{+}&=& - p^{+} \left\{2\eta G_3\right\},
\label{1-1} \\
J_{11}^{+}&=&p^{+} \left\{2 G_1+2\eta G_3\right\}
\label{11}
\end{eqnarray}
where $J_{\lambda \lambda '}^{+}\equiv
J_{\lambda \lambda '}^{0} + J_{\lambda \lambda '}^{3}$.
It is easy to show that they satisfy the so-called angular condition
\begin{equation}
(1+2\eta)J_{11}^{+}+J_{1-1}^{+}-2\sqrt{2\eta}J_{10}^{+}
-J_{00}^{+}=0
\label{angular_c}
\end{equation}
and thus there are only three independent helicity amplitudes between
the $J_{11}^{+}$, $J_{1-1}^{+}$, $J_{10}^{+}$ and $J_{00}^{+}$
\cite{GrachKondratyuk,Coester}.

Alternatively the angular condition (\ref{angular_c}) teaches us that even 
at pQCD extreme there appears (through dimensionless ratio
$\eta=\frac{Q^2}{4M^2}$) an additional scale parameter $4M^2$, apart from the
pQCD parameter $\Lambda_{\rm QCD}^2$.

\subsection{The deuteron form factors}

The charge, $G_C(Q^2)$, magnetic, $G_M(Q^2)$, and quadruple,
$G_Q(Q^2)$, form factors are connected with the form factors $G_1(Q^2)$,
$G_2(Q^2)$ and $G_3(Q^2)$ as follows
\begin{eqnarray}
G_Q &=& G_1-G_2+(1+\eta)G_3,\nonumber \\
G_C &=& G_1+\frac{2}{3}\eta G_Q,\nonumber \\
G_M &=& G_2.
\label{9}
\end{eqnarray}
Using (\ref{00})-(\ref{11}) one expresses $G_C(Q^2)$, $G_M(Q^2)$ and
$G_Q(Q^2)$ in terms of any three helicity amplitudes
$J^{+}_{\lambda \lambda'}$, for example
\begin{eqnarray}
G_C &=& \frac{1}{2\cP(2\eta+1)}\left[
\frac{3-2\eta}{6}J^{+}_{00}+
\frac{8}{3}\sqrt{\frac{\eta}2}J^{+}_{10}+\frac{2\eta-1}{3}J^{+}_{1-1}
\right],\\
\nonumber
G_M &=& \frac{1}{2\cP(2\eta+1)}\left[
J^{+}_{00}+\frac{(2\eta-1)}{\sqrt{2\eta}}
J^{+}_{10}-J^{+}_{1-1}\right],\\
\nonumber
G_Q &=& \frac{1}{2\cP(2\eta+1)}\left[
-\frac12J^{+}_{00}+
\sqrt{\frac{1}{2\eta}}J^{+}_{10}-
\frac{\eta+1}{2\eta}J^{+}_{1-1}\right].
\label{BH_set}
\end{eqnarray}
In turn, the $A(Q^2)$ and $B(Q^2)$ structure functions and the
$t_{2i}(\theta,Q^2)$ polarizations read 
\begin{eqnarray}
A &=& G_{C}^{2}+\frac{2}{3}\eta~G_{M}^{2}+\frac{8}{9}\eta^{2}G_{Q}^{2},
\label{A(Q2)}\\
B &=& \frac{4}{3}\eta~(1+\eta)G_M^{2},
\label{B(Q2)}\\
t_{20} &=&
-\frac{1}{\sqrt{2}\,{\cal S}}
\left\{
\frac{8}{9}\eta^{2}~G_Q^{2}+\frac{8}{3}\eta~G_C~G_Q+
\right.
\nonumber \\
&& \left. +
\frac{2}{3}\eta~G_M^{2}
\left[\frac{1}{2}+(1+\eta){\rm tg}^{2}\frac{\theta}{2}
\right]
\right\},
\label{t20}\\
t_{21} &=& \frac{2}{\sqrt{3}\,{\cal S}\cos\frac{\theta}{2}}
\eta\left(
\eta+\eta^2\sin^2\frac{\theta}{2}  \right)^{\frac12}G_M G_Q,
\label{t21} \\
t_{22} &=& -\frac{1}{2\sqrt{3}\,{\cal S}}G_M^2,
\label{t22}
\end{eqnarray}
where
${\cal S}=A^{2}+B^{2}{\rm tg}^{2}\frac{\theta}{2}$.

\subsection{pQCD predictions for the helicity amplitudes}
From pQCD arguments one gets very simple rules to determine the power
behavior of the helicity amplitudes $J_{00}^{+}$, $J_{10}^{+}$ and
$J_{1-1}^{+}$ \cite{VainshteinZakharov,CarlsonGross}. For example, it
follows that the amplitude $J_{00}^{+}$ is a leading amplitude with
an asymptotic fall off 
\begin{equation}
J_{00}^{+} \sim \left( \frac{\Lambda_{\rm QCD}}{Q} \right)^{10}
\label{17}
\end{equation}
up to logarithmic corrections.
It was also argued that in LCF the helicity flip amplitudes
$J_{10}^{+}$ and $J_{1-1}^{+}$ are suppressed as \cite{BH2}
\begin{equation}
\frac{J_{10}^{+}}{J_{00}^{+}} \sim \frac{\Lambda_{\rm QCD}}{Q},\
\frac{J_{1-1}^{+}}{J_{00}^{+}} \sim
\left(\frac{\Lambda_{\rm QCD}}{Q}\right) ^{2}.
\label{18}
\end{equation}
Similar considerations give that
\begin{equation}
J^{+}_{11} \sim \left(\frac{M}{\Lambda_{\rm QCD}}\right)^2
\left(\frac{\Lambda_{\rm QCD}}{Q}\right)^{2}J^+_{00} ,
\label{19}
\end{equation}
which agrees with the angular condition (\ref{angular_c}), but only at
extremely high $Q^2$, $Q^2\gg 4 M^2$.

In our analysis, following arguments of Ref.~\cite{BH2,KS1}, we consider
the set $J_{00}^{+}$, $J_{10}^{+}$ and $J_{1-1}^{+}$ as main
amplitudes, where the behavior (\ref{17}),(\ref{18}) is regulated in the
intermediate region only by the $\rm \Lambda_{\rm QCD}$. In turn, the 
amplitude $J_{11}^{+}$ must be determined from the angular condition
(\ref{angular_c}) and thus it depends on the two scale parameters, 
$\Lambda_{\rm QCD}$ and $4M^2$. 

\section{Parameterization of helicity transition amplitudes}

Following the idea of reduced nuclear amplitudes in QCD
\cite{BrodskyChertok}, we define the reduced helicity transition
amplitudes $g_{00}$, $g_{0+}$ and $g_{+-}$ as follows:
\begin{equation}
\frac1{2{\cal P}}J_{\lambda,\lambda'}^{+}(Q^{2})=
G^{2}\left(\frac{1}{4}Q^2\right) g_{\lambda,\lambda'}(Q^{2}),
\label{20}
\end{equation}
where $G(Q^{2})$ is a three-quark-cluster (nucleon) form factor. For the
$G(Q^{2})$ we assume dipole behavior
$G(Q^{2})=\left[1+\frac{Q^{2}}{\mu^2}\right]^{-2}$, but with 
the parameter $\mu^2$ is different  from
that for a free nucleon $0.71\ {\rm (GeV/c)}^2$.

We consider separately two kinematical regions, large $Q^{2}$ region,
$Q^{2} > Q^2_{\rm QCD}$, and low $Q^{2}$ region, $Q^{2}<Q^2_{\rm QCD}$. The parameter
$Q^2_{\rm QCD}$ is expected to be of order of few $\rm (GeV/c)^2$. Its exact
value will be determined from a fit to experimental data.

pQCD predicts that at asymptoticaly high $Q^{2}$ the reduced transition 
amplitudes behave as follows
\begin{eqnarray}
g^{(\rm asy)}_{00} &=&
\frac{N_1}{Q^{2}}\phi(Q^{2}), \
g^{(\rm asy)}_{0+}=\frac{N_2}{Q^{3}}\phi(Q^{2}),
\nonumber \\
g^{(\rm asy)}_{+-} &=& \frac{N_3}{Q^{4}}\phi(Q^{2}).
\label{21}
\end{eqnarray}
In (\ref{21}) the factor $\phi(Q^{2})$ takes into account logarithmic
corrections
\begin{equation}
\phi(Q^{2})=\frac{[\alpha_{s}(Q^{2})]^{5}}{[\alpha_{s}(Q^{2}/4)]^{4}}
\frac{(\lg Q^{2}/\Lambda_{\rm QCD}^{2})^{\gamma^{d}}}{
[\lg Q^{2}/(4\Lambda_{\rm QCD}^{2})]^{\gamma^{N}}}
\label{22}
\end{equation}
and $\gamma^{d}$ and $\gamma^{N}$ are leadings anomalous dimensions
for the deuteron and the nucleon, respectively,
\begin{equation}
\gamma^{d}=\frac{6C_{F}}{5\beta}, \gamma^{N}=\frac{C_F}{2\beta},
\label{23}
\end{equation}
where $C_{F}=(n_c^2-1)/(2n_c)$, $\beta=11-\frac{2}{8}n_f$,
$n_c=3$  is number of quark colors and $n_f=2$ is the number
of flavors, $\alpha_{s}(Q^{2})=\frac{4\pi}{\beta~\ln Q^2/\Lambda_{\rm QCD}^2}$ is
the running quark-gluon coupling \cite{BJL}.

At $Q^{2}\le Q_{\rm QCD}^2$ the following
parameterization for the reduced amplitudes is assumed
\begin{eqnarray}
\tilde g_{00} &=& \sum_{n=1}^{N}\frac{a_n}{Q^{2}+\alpha_{n}^{2}},\
\tilde g_{0+} = Q~\sum_{n=1}^{N}\frac{b_n}{Q+\alpha_{n}},\\
\nonumber
\tilde g_{+-} &=& Q^{2}\sum_{n=1}^{N}\frac{c_n}{Q^{2}+\alpha_{n}^{2}},
\label{24}
\end{eqnarray}
where $\alpha_{n}^2=\left[\alpha_{0} + (n-1) m_0\right]^2$.
One imposes the constrains (see, e.q., \cite{KS2}):
\begin{eqnarray}
\sum_{n=1}^{N}\frac{a_n}{\alpha_{n}^2} &=& 1,\
\sum_{n=1}^{N}\frac{b_n}{\alpha_{n}^2} = \frac{2-\mu_{d}}{2~\sqrt{2}~M},\nonumber \\
\sum_{n=1}^{N}\frac{c_n}{\alpha_{n}^2} &=& \frac{1-\mu_{d}-Q_{d}}{2M^{2}}
\label{25}
\end{eqnarray}
on the coefficients $a_n$, $b_n$
and $c_n$ to demand form factor normalization at $Q^2=0$. In (\ref{25})
$\mu_{d}=0.857406~M/m_p$ is the deuteron magnetic moment in
``deuteron magnetons'' and $Q_{d}=25.84$ is the deuteron quadrupole
momentum in $M^2/e$.

The coefficients $N_1$, $N_2$ and $N_3$ appearing in (\ref{21}) cannot be 
calculated from pQCD and we determine them from the smooth connection of the 
two parametrizations at the point $Q^{2}=Q_{\rm QCD}^{2}$
\begin{eqnarray}
g^{(\rm asy)}_{ij} &=& \tilde g _{ij},\nonumber \\
\frac{d^{2}~g^{(\rm asy)}_{00}}{d Q^{2}} &=&
\frac{d^{2}~\tilde g _{00}}{d
Q^{2}},\\ \nonumber
\frac{d~g^{(\rm asy)}_{0+}}{d Q} &=&
\frac{d~\tilde{g}_{0+}}{d Q},\nonumber \\
\frac{d^{2}~g^{(\rm asy)}_{+-}}{d Q^{2}} &=&
\frac{d^{2}~\tilde{g}_{+-}}{d Q^{2}}.
\label{26}
\end{eqnarray}

\section{Data base and fitting parameters}

In out fit we use the following data:
for $A(Q^2)$ from 
\cite{Alexa,AbbottA,Arnold,Buchanan,Cramer,Drikley,Elias,Gaster,Plachkov},
for $B(Q^2)$ from  \cite{Bosted,Buchanan,Cramer,Aufert,Simon}, 
and for  $t_{20}$ from
\cite{Abbottt20,Bouwhuis,Dmitriev,FerroL,Garcon,Schulze,Voitsekhovskii}
and for $t_{22}$, $t_{21} $ from \cite{Abbottt20,Garcon}. But it must be noted 
that due to large errors $t_{22}$, $t_{21}$ data are practically not 
informative for the fit. Data for $A$ from \cite{Simon} are in sharp contradiction
with all the wold data set and therefore were omitted from the fit. We also do not 
include in the data base results from \cite{Akimov} because they have large errors
and practically do not change the obtained results. Data \cite{Elias}
are somewhat lower that the world data, but due to large their uncertainties the fit is
insensitive to this data. Data for the $B$ structure
function were normalized, if necessary, to our convention of the magnetic form factor.

$Q_{\rm QCD}^2$ was  considered as a parameter of the model, but the QCD cutoff
parameter $\Lambda_{\rm QCD}$ was fixed at 200 MeV. In (\ref{25}) we chose $N=4$, so 
that the the model has 10 independent fit parameters. The fit parameters summarized 
in Table 1 were obtained with $\chi^2=399$ for 200 data points.

\section{Discussion and summary}

Figs.~1 and 2 display comparison of the model with $A(Q^2)$ data. 
To show how far experimental data are from pQCD results in preasymptotic region
we continue asymptotic behavior given by (\ref{21}) to  region
lower than $ Q_{\rm QCD}^2$ (dotted line on Fig.~2 and Figs.~3-7).
One concludes that for $A(Q^2)$ pQCD works from $Q^2$-region between 1 and 2 $({\rm GeV/c})^2$.  
Comparison with data for $B(Q^2)$ and the polarization observables are 
given on Figs.~3 --- 5. One sees that for the magnetic form factor pQCD should 
start from $Q^2$ between 2 
and 3 $\rm GeV/c$, but for the polarization observable it should start somewhat earlier,
near 2 $({\rm GeV/c})^2$ .

Fig. 7 shows results for for the charge and quadrupole form factors, 
$G_c(Q^2)$ and $G_q(Q^2)$.

In summary, we give a parameterization of the deuteron form factors up to 
$Q^2=6\ {\rm (GeV/c)^2}$. Asymptotic behavior of the form factors is dictated
by quark counting rules and pQCD helicity rules and 
therefore one may hope that this 
parameterization can be extrapolated for higher transferred momentum. 
For example, the model predicts behavior of 
the magnetic structure function, $B(Q^2)$,  at $Q^2\ge 2.5\ ({\rm GeV/c})^2$
which can be studied in future experiments.

\section*{Acknowledgments}
We thank M.~Gar\c{c}on, G.G.~Petratos and E.A.~Strokovsky for number of helpful
discussions. One of the authors (A.K.~K.) acknowledge the hospitality of RCNP,
where this work was carried out with a Center of Excellence grant from the
Ministry of Education, Culture, Sports, Science and Technology (Monbu-Kagaku-sho), Japan.


\newpage
\begin{table}
\caption{Parameters of the model.}
\begin{center}
\begin{tabular}{|| l | l  l ||}
\hline 
$ \alpha_0$ & 0.2635 & GeV \\
\hline
$ m_0 $     & 0.36864 & GeV\\
\hline
$a_1$       & $0.5373059\cdot 10^{-1}$ &$\rm GeV^{-2}$ \\
\hline
$a_2$       & 0.2836982               &$\rm GeV^{-2}$\\
\hline
$a_3$       & from eqs. (28) &\\
\hline
$a_4$       & from eqs. (27) &\\
\hline
$b_1$       & $-9.898079\cdot 10^{-1}$ & \\
\hline
$b_2$       & \ \  1.6950028  &\\
\hline
$b_3$       & from eqs. (28) &\\
\hline
$b_4$       & from eqs. (27) &\\
\hline
$c_1$       & -0.1074219 & \\
\hline
$c_2$       & $-0.7075449\cdot 10^{-1}$  &\\
\hline
$c_3$       & from eqs. (28) &\\
\hline
$c_4$       & from eqs. (27) &\\
\hline
$Q^2_{\rm QCD}$ & 3.50120 & $\rm GeV^2$ \\
\hline
$\mu^2$ & 0.50959 & $\rm GeV^2$ \\
\hline 
\end{tabular}
\end{center}
\end{table}

\mbox{}

\newpage
\begin{center}{\bf \large Figure captions}
\end{center}

\mbox{}\\
{\bf Figure 1.} 
Comparison of the model fit with data for $A(Q^2)$ at
$Q^2\ \le\ 1\ {\rm (GeV/c)^2 }$.
\mbox{}\\
{\bf Figure 2.} 
Comparison of the model fit (solid line) with data for $A(Q^2)$ at
$Q^2\ \ge\ 1\ {\rm (GeV/c)^2 }$. The dotted line is asymptotic
behavior given by (\ref{21}) extrapolated to lower transfer momentum.
\mbox{}\\
{\bf Figure 3.} 
Comparison of the model fit with data for $B(Q^2)$. 
The lines are the same as at Fig.~2.
\mbox{}\\
{\bf Figure 4.} 
Comparison of the model fit with data for $t_{20}$. 
The lines are the same as at Fig.~2.
\mbox{}\\
{\bf Figure 5.} 
Comparison of the model fit with data for $t_{21}$. 
The lines are the same as at Fig.~2.
\mbox{}\\
{\bf Figure 6.} 
Comparison of the model fit with data for $t_{22}$. 
The lines are the same as at Fig.~2.
\mbox{}\\
{\bf Figure 7.} 
Comparison of the model fit with data for $|G_c(Q^2)|$
(upper figure) and  $G_q(Q^2)$ (lower figure). Data are from
\cite{Abbott_par}. 
For the last two points the two solutions (filled and open points)
are shown, see \cite{Abbott_par}.
The lines are the same as at Fig.~2.

\newpage
\includegraphics{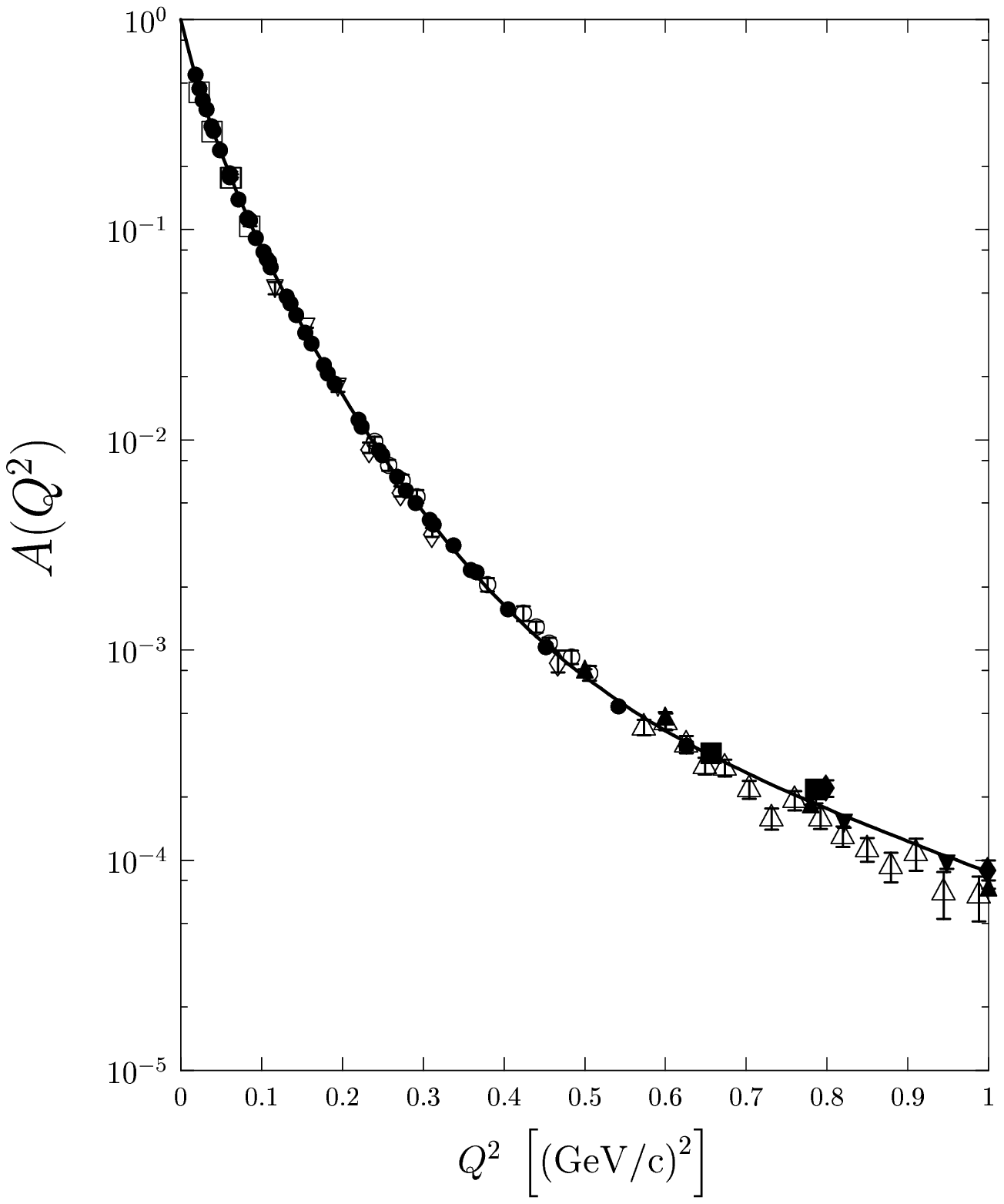}\includegraphics{a1.ps}
\mbox{}
\vfill \begin{center}{\LARGE Fig.~1.}
\end{center}

\newpage
\includegraphics{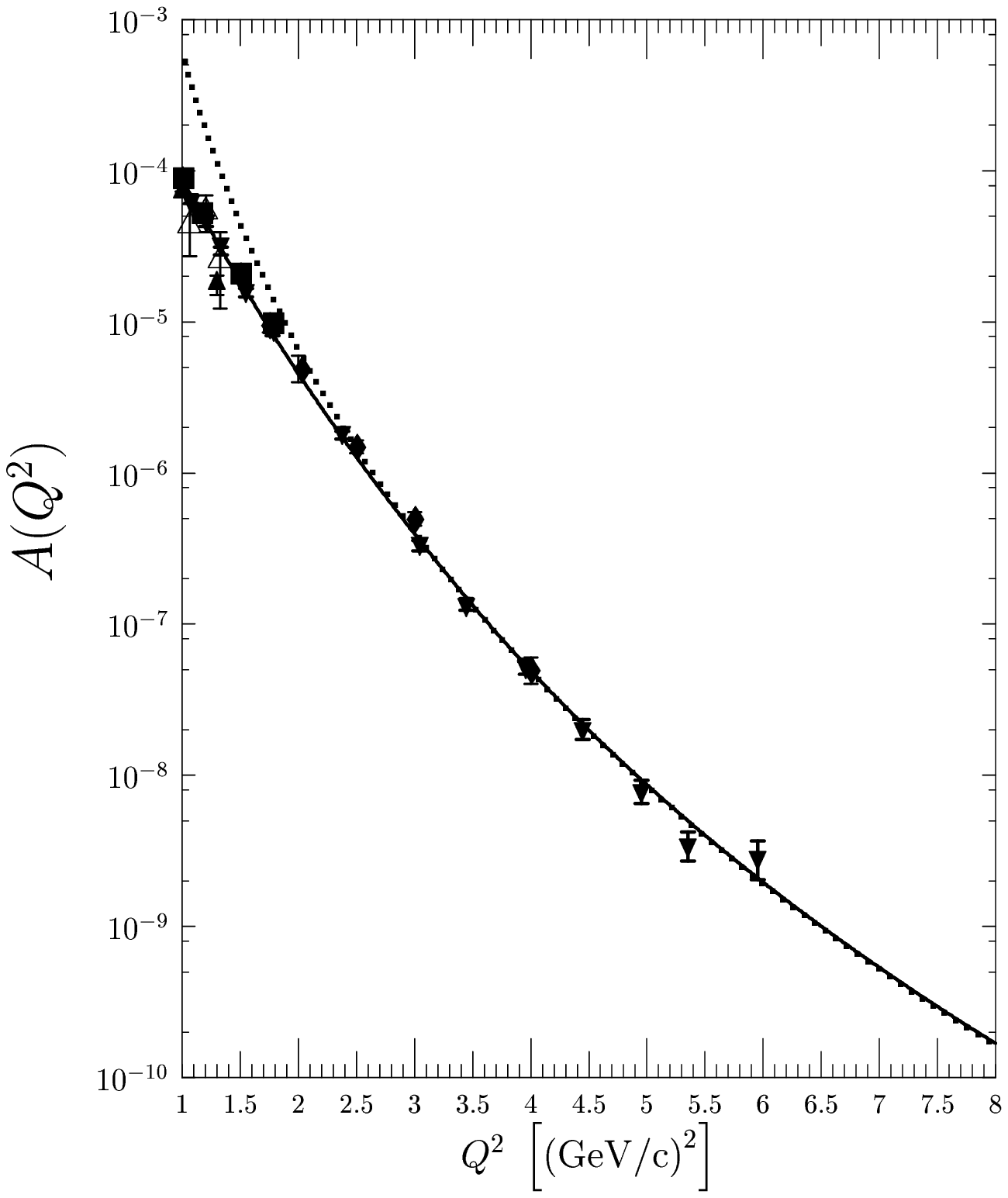}
\mbox{}
\vfill \begin{center}{\LARGE Fig.~2.}\end{center}

\newpage
\includegraphics{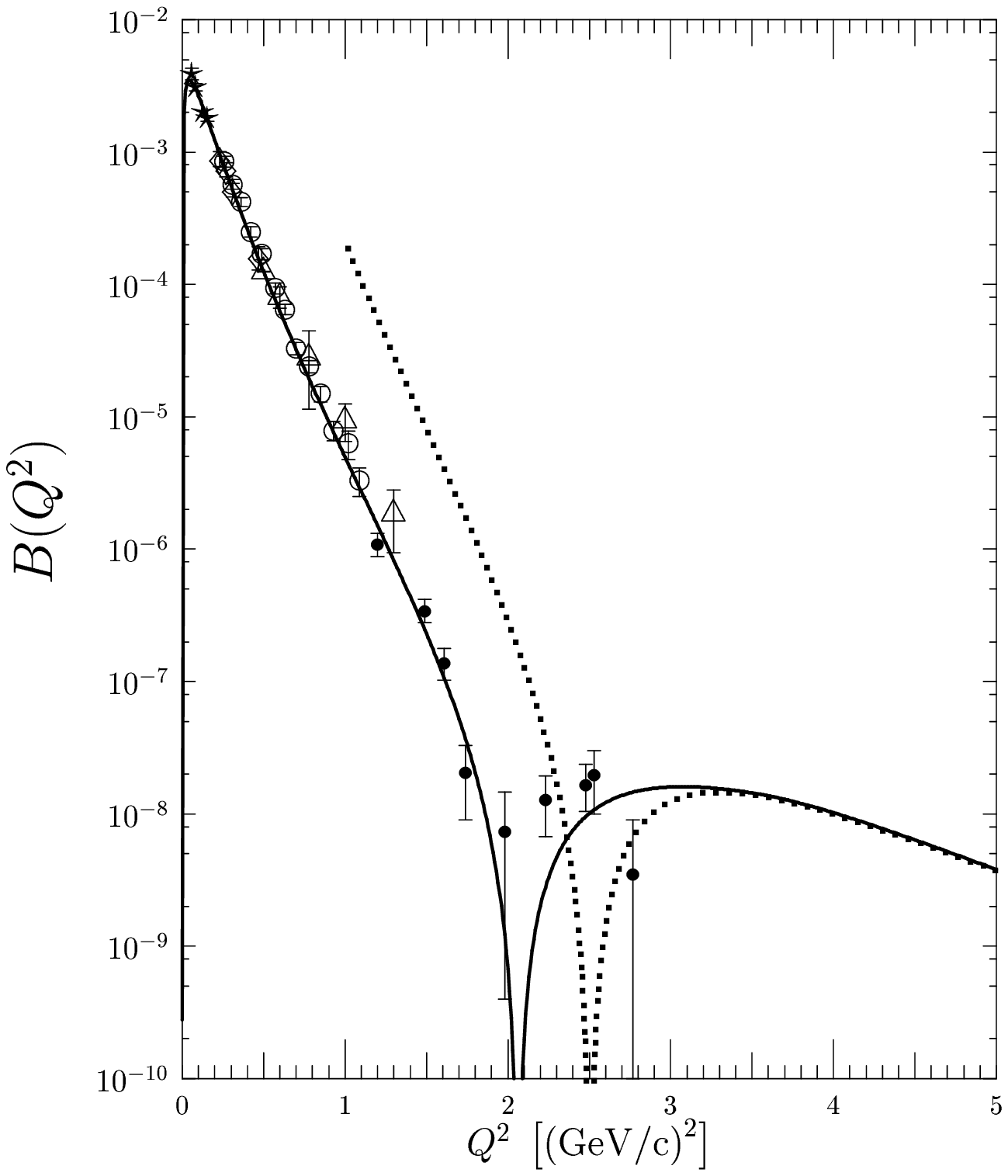}
\mbox{}
\vfill \begin{center}{\LARGE Fig.~3.}\end{center}

\newpage
\includegraphics{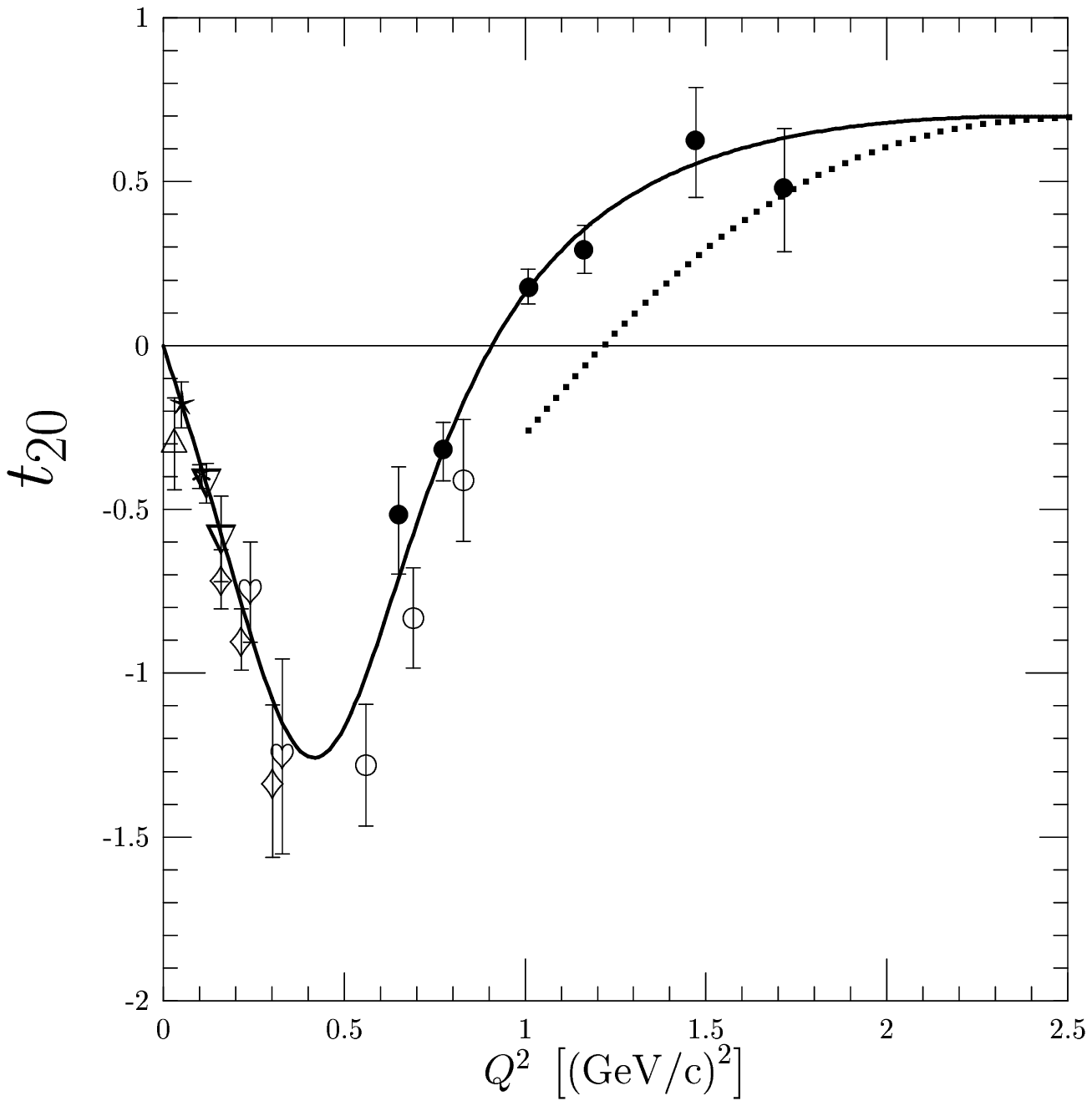}
\mbox{}
\vfill \begin{center}{\LARGE Fig.~4.}\end{center}

\newpage
\includegraphics{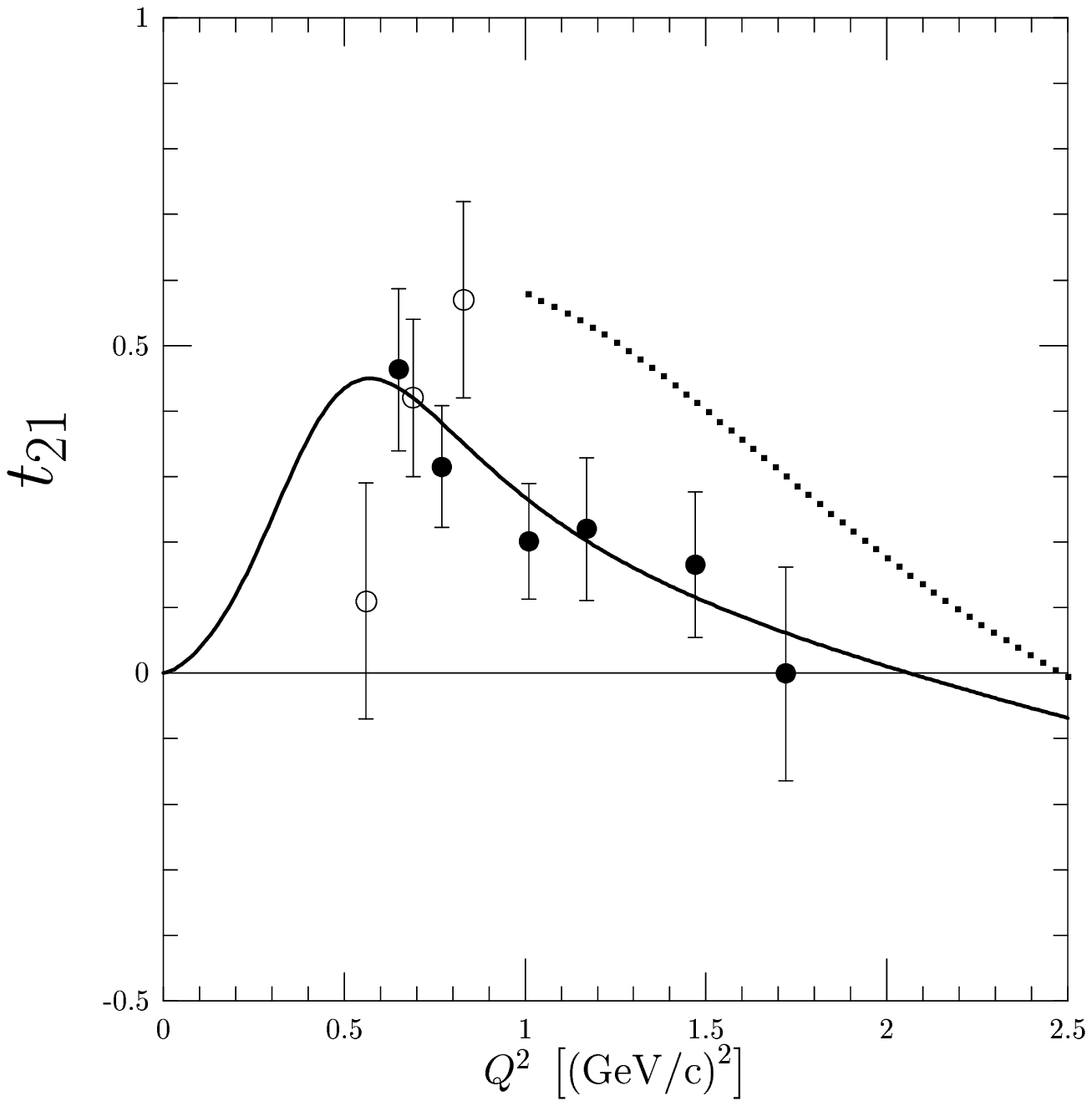}
\mbox{}
\vfill \begin{center}{\LARGE Fig.~5}\end{center}
\newpage
\mbox{}
\includegraphics{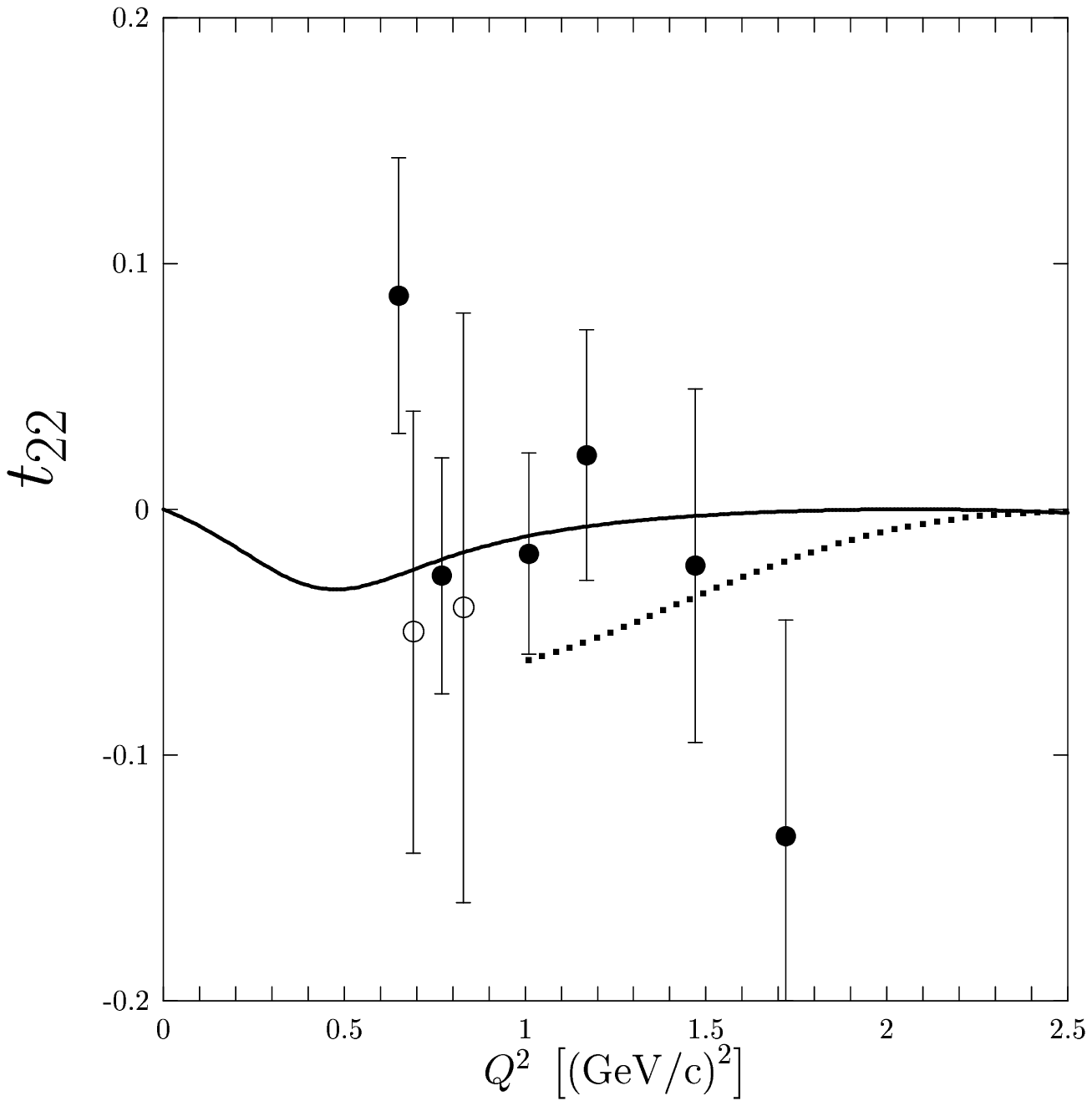}
\mbox{}
\vfill \begin{center}{\LARGE Fig.~6}\end{center}

\newpage
\includegraphics{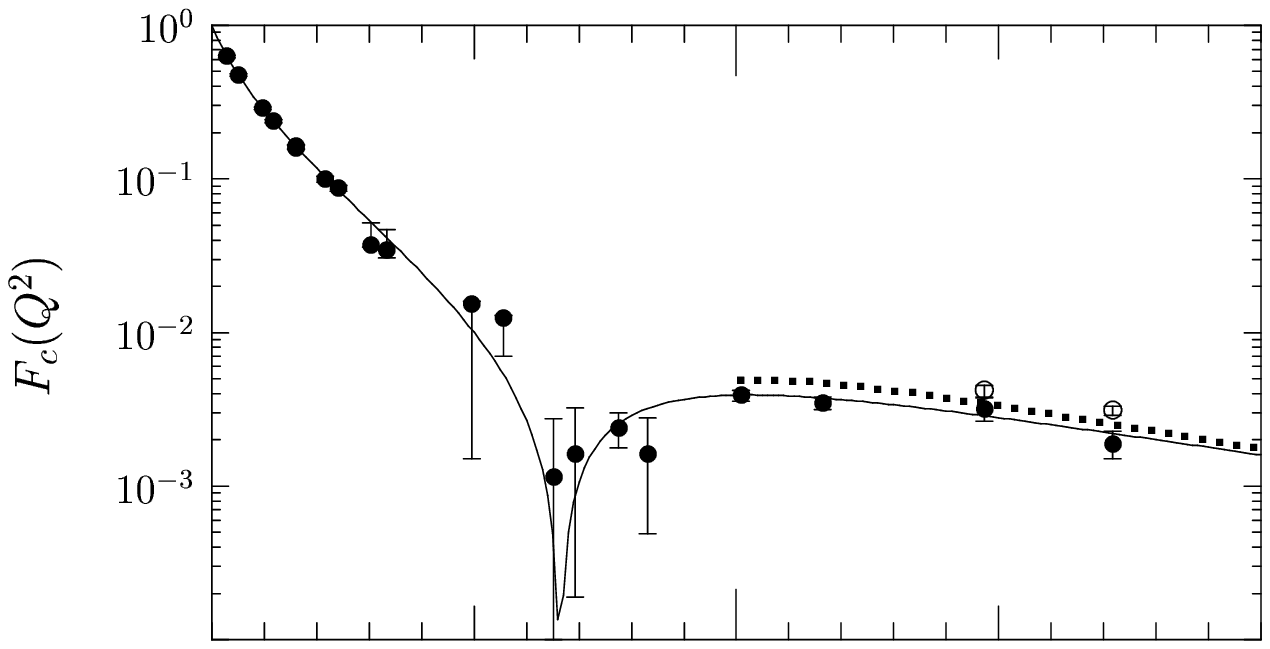}
\mbox{}

\includegraphics{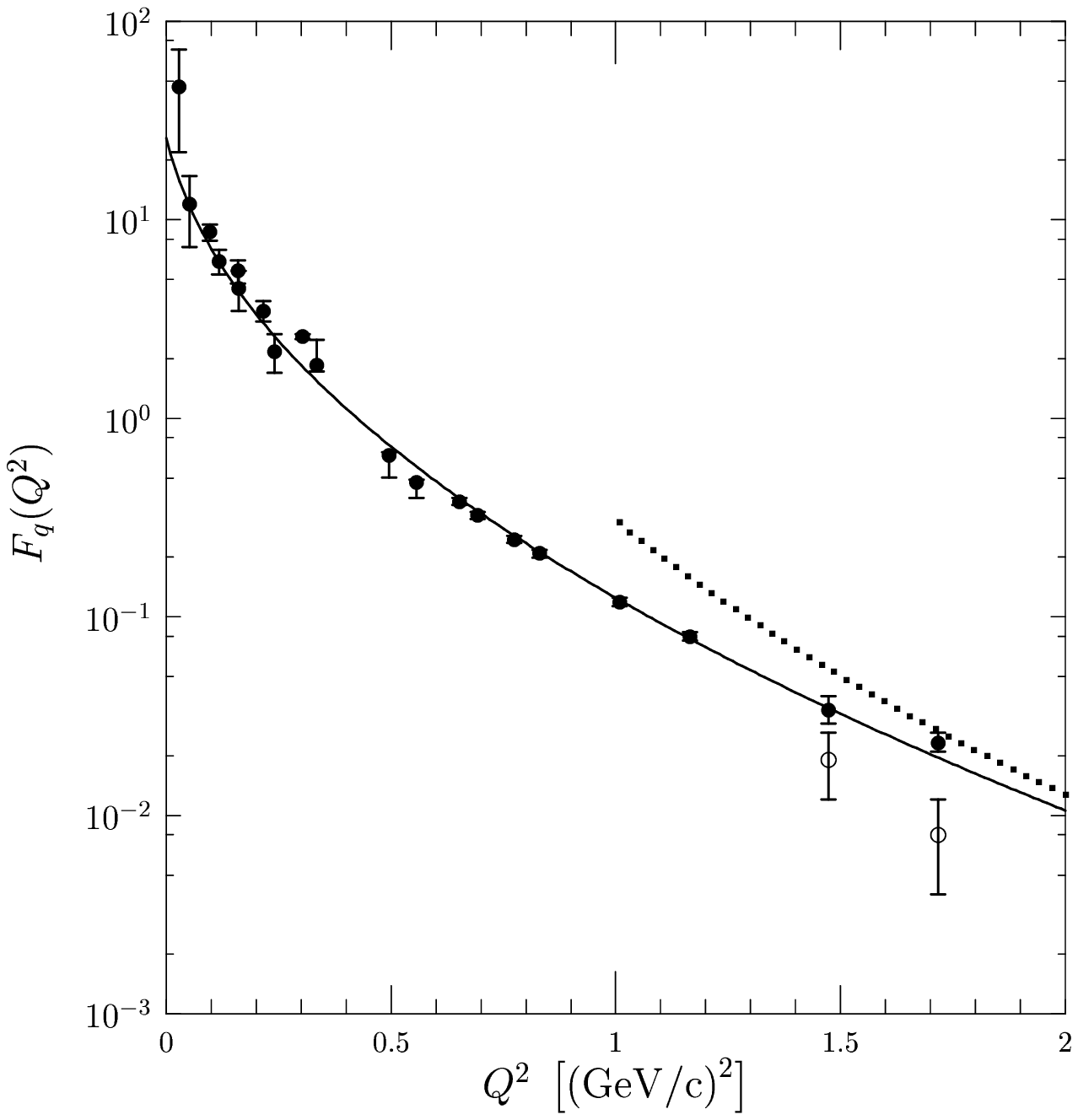}
\mbox{}
\vfill \begin{center}{\LARGE Fig.~7.}\end{center}

\end{document}